\documentclass[12pt]{iopart}
\usepackage{iopams}
\usepackage{xcolor}
\usepackage{graphicx}
\usepackage{setstack}
\usepackage{braket}
\begin{document}
	\title{Satellite-based links for Quantum Key Distribution: beam effects and weather dependence}
	\author{Carlo Liorni$^1$, Hermann Kampermann$^1$, Dagmar Bru\ss$^1$}
	\address{$^1$Heinrich Heine Universit\"at, Institut f\"ur Theoretische Physik III, Universit\"atsstra\ss e 1, 40235, D\"usseldorf, Germany.}
	
	\begin{abstract}
		The establishment of a world-wide quantum communication network relies on the synergistic integration of satellite-based links and fiber-based networks. The first are helpful for long-distance communication, as the photon losses introduced by the optical fibers are too detrimental for lengths greater than about 200 km. This work aims at giving, on the one hand, a comprehensive and fundamental model for the losses suffered by the quantum signals during the propagation along an atmospheric free-space link. On the other hand, a performance analysis of different Quantum Key Distribution (QKD) implementations is performed, including finite-key effects, focusing on different interesting practical scenarios. The specific approach that we chose allows to precisely model the contribution due to different weather conditions, paving the way towards more accurate feasibility studies of satellite-based QKD missions. \\

	\end{abstract}
	\noindent{\it Keywords\/}: Satellite links, Quantum Key Distribution, Atmospheric effects \\
	\maketitle
	
	\section{Introduction}
	\label{intro}
	
	Quantum Key Distribution (QKD) and Quantum Communications in general have the potential to revolutionise the way we communicate confidential information over the internet. The natural carriers for quantum information are photons, that are already widely used in classical networks of optical fibers to achieve high communication rates. Unfortunately, even though enormous improvements have been obtained in the last years \cite{exp1,exp2}, scaling quantum communication protocols over long distances is very challenging, due to the losses experienced during the propagation inside the optical fibers. Several schemes for the realization of quantum repeaters have been proposed in recent years, that could allow to bridge long distances and naturally be implemented inside a quantum communication network \cite{rep1,rep2,rep3,rep4,rep5}.
	Considering the important technological hurdles that quantum repeaters should overcome before becoming useful, satellite-based free-space links look like the most practical way to achieve long-distance QKD in the short term \cite{jennewein2}. They can take advantage of the satellite technology and the optical communication methods developed in the last decades in the classical case. Various feasibility studies had addressed this topic in the last twenty years \cite{Bedington,jennewein1,jennewein2,bonato} and several experiments have definitely proved that the technology involved is ready for deployment \cite{Liao,pan2,pan3,pan4,takenaka}. 
	
	Optical satellite-based links have the important drawback of being strongly dependent on the weather conditions \cite{weather1,weather2,weather3,weather4}. The presence of turbulent eddies and scattering particles like haze or fog generates random fluctuations of the relative permittivity of the air, on different length- and time-scales. This phenomenon affects the light propagation in a complicated way, inducing random deviations and deformations of any optical beam sent through the atmosphere. It results in reduced transmittance, because of geometrical losses due to the finite collection aperture, and random modifications of the phase front. A comprehensive model of these effects is then necessary, in order to precisely evaluate the performances of the link when used for quantum communication protocols. 
	
	In this work we generalize the approach proposed in \cite{vogel1,vogel2} to satellite-based links and we evaluate their losses in several practical cases, under different weather conditions. This information is then used to assess the performances of the link in terms of the achievable key rates using different implementations of QKD. The case of \emph{Low Earth Orbit} (LEO) satellites is addressed, assuming different payloads and sizes of the optical elements.  
	
	This work is organized as follows. In Sec. \ref{freeintro} we introduce the problem of free-space optical links and an analytical method to study them. The discussion continues in \ref{free}.
	In Sec. \ref{modelresults} a detailed description of the model used to simulate the satellite-based link is presented. Then, the main results are shown and discussed, together with pros and cons of our approach. In Sec. \ref{perf} we use the analysis of the transmittance of the channel conducted in the previous section to study the performances of different QKD implementations, in some interesting real-life scenarios. The analysis concerning the use of smaller and more affordable satellites is performed in Sec. \ref{cube}. Finally, the results are summarized and discussed in Sec. \ref{discussion}. The appendix starts with a recap of the results of \cite{vogel1,vogel2} (\ref{free}) and their application to the problem at hand (\ref{example}). Then two models for the estimation of the stray light satellite links \cite{bonato,miao} are presented in \ref{envphot}. \ref{rates} is devoted to the definition of the QKD protocols we use in Sec. \ref{perf} and the expression of the correspondent key rates. In \ref{parameters} we report the parameters chosen for the simulations and we discuss their pertinence.

	\section{Free-space optical links and the Elliptic Beam Approximation}
	\label{freeintro}
	The problem that we address in the first part of the work is the following. A Gaussian beam is sent, either from an orbiting transmitter or from a ground station, through a non-uniform link partially inside the atmosphere and partially in vacuum. We are interested in the transmittance of the received beam through a circular aperture of radius $a$ (the receiving telescope)
	\begin{equation}
	\label{eta}
	\eta=\int_{|\brho|^2=a^2} d^2\brho  \ |u(\brho,L)|^2 \ ,
	\end{equation}
	which is a random variable, because of the intrinsic randomness of the fluctuations in the medium. Here $u(\brho,L)$ is the beam envelope at the receiver plane (at distance $L$ from the transmitter, with $\brho$ the position in the transverse plane). 
	
\begin{figure}[ht]
		\includegraphics[width=0.35\textwidth]{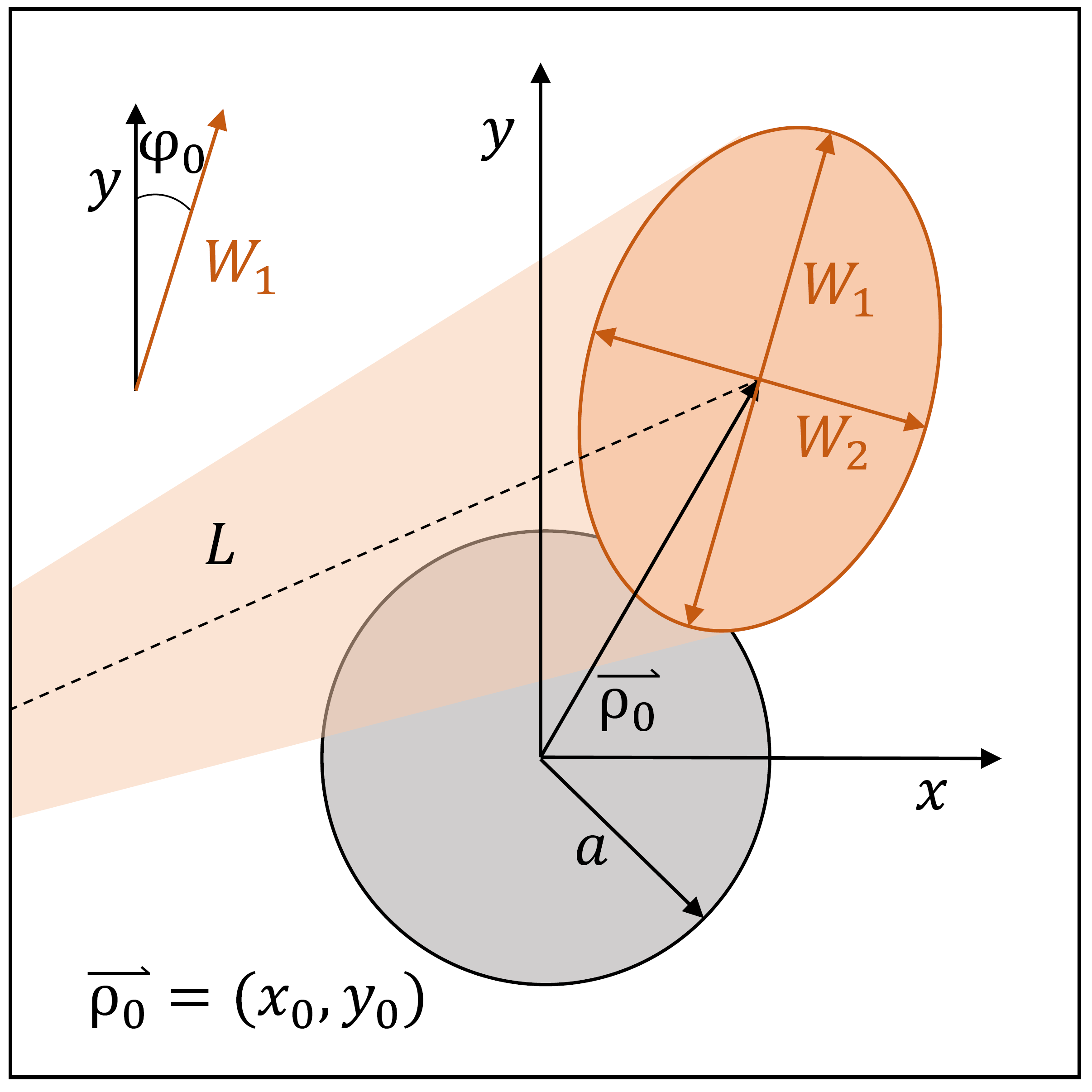}
		\centering
		\caption{\label{fig:beam} Schematic representation of the received beam and receiving aperture. $L$ is the length along the propagation direction, $a$ the radius of the receiving aperture, $\brho_0=(x_0,y_0)$ is the beam-centroid position, $W_1$ and $W_2$ the two axes of the elliptical profile, $\varphi_0$ the angle of orientation of the ellipse.}
	\end{figure}

	The so-called \emph{Elliptic Beam Approximation} \cite{vogel1} greatly simplifies the analysis: the atmosphere is assumed to generate only
	\begin{itemize}
		\item deflection of the beam as a whole (\emph{Beam Wandering})
		\item elliptic deformations of the beam profile 
		\item extinction losses due to back-scattering and absorption.
	\end{itemize} 
	In this case the state of the beam at the receiver plane is completely described by the vector of parameters (refer to Fig \ref{fig:beam})
	\begin{equation}
	\label{paramv}
	{\bf v} = (x_0,y_0,W_1, W_2, \varphi_0) \ ,
	\end{equation}
	representing the beam-centroid coordinates, the principal semi-axes of the elliptic profile and the angle of orientation of the ellipse. The transmittance is then a function of these beam parameters and the radius of the receiving aperture.
	
	The fluctuations of the relative permittivity of the atmospheric air can be statistically modeled \cite{kolmogorov,turb43,turb44,turb45,scatt19,scatt20,scatt21,scatt22,scatt23}. The probability distribution of the parameters in Eq.~(\ref{paramv}) can then be analytically estimated, as shown in \cite{vogel1,vogel2}. A brief recap of the derivation and the main results is presented in \ref{free}.
	This allows, through random sampling, to obtain the Probability Distribution of the Transmittance (PDT), an important figure of merit for fluctuating links. This approach gives no information about the phase of the wavefront, but this is not a problem when phase-insensitive measurements are considered (e.g., the BB-84 QKD protocol that we analyze in Sec.~\ref{perf}).

	\section{Satellite-based links: model and results}
	\label{modelresults}
	
\begin{figure}[ht]
		\includegraphics[width=0.5\textwidth]{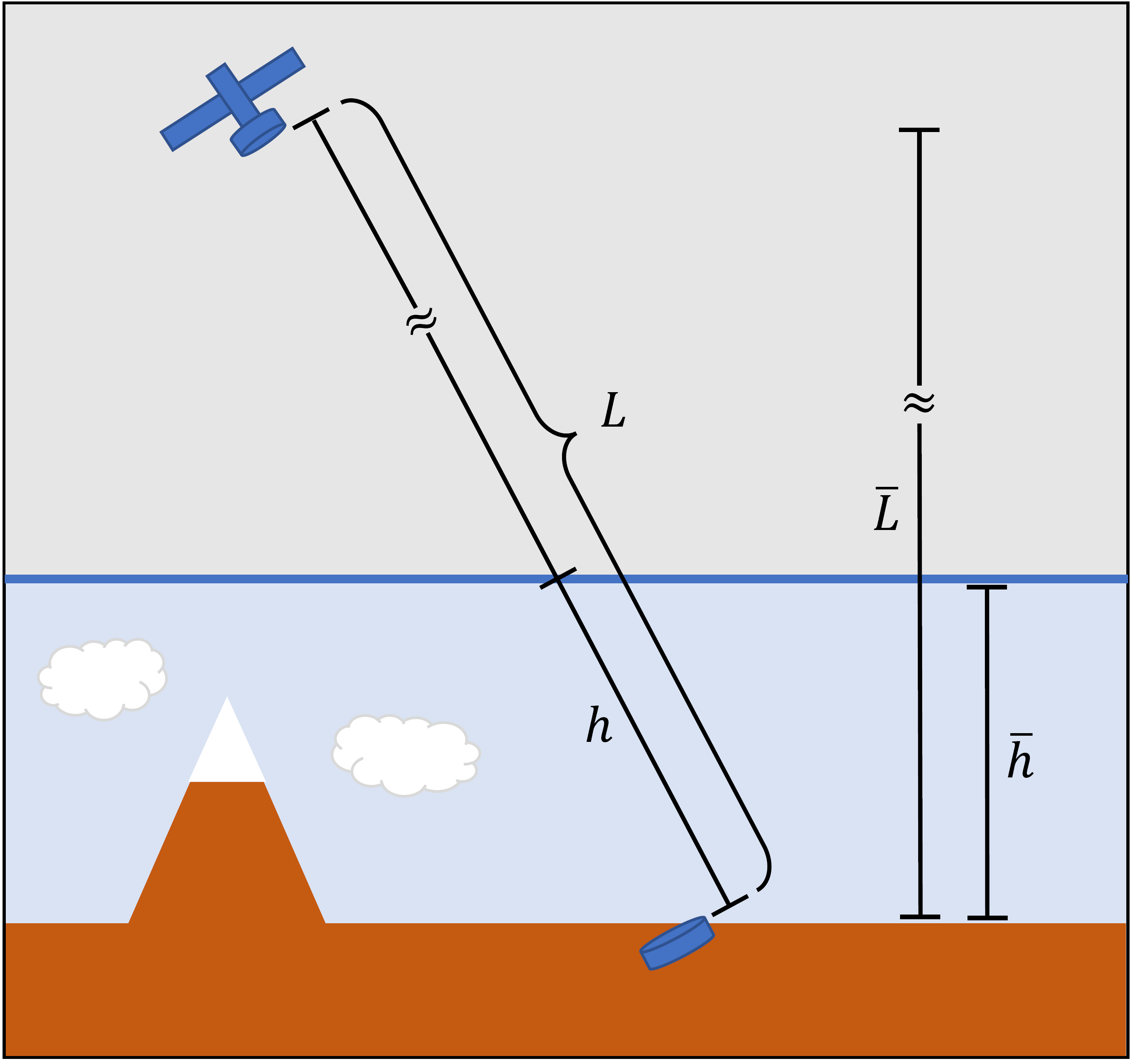}
		\centering
		\caption{\label{fig:atmo} The non-uniform free-space link between the satellite and the ground station is depicted here (not in scale). The main parameters shown are the thickness of the atmosphere $\bar{h}$, the height of the satellite $\bar{L}$, the total distance between sender and receiver $L$ and the length of the propagation inside the atmosphere $h$.}
	\end{figure}

	The atmosphere can in general be divided into several layers, depending on the properties of different physical parameters, like density of the air, pressure, temperature, density of ionized particles, and so on. This structure is site-dependent, especially regarding the thickness of the different layers. For this reason, in this work we assume a simplified version of a satellite-based optical link: a uniform atmosphere up to a certain altitude $\bar{h}$, then vacuum all the way up to the satellite (at altitude $\bar{L}$), as pictured in Fig.\ref{fig:atmo}. Instead of a continuum of values describing the physical quantities as a function of the altitude, we now have only two parameters, namely the value of the quantity inside the uniform atmosphere and the effective thickness $\bar{h}$. This is likely to be a good approximation, because the atmospheric effects are prominent only in the first 10 to 20 km from the ground, while usual orbit height for LEO satellites are above 400 km. For the remainder of the paper we choose a minimum altitude of the satellite $\bar{L}=500$ km, achieved exactly above the ground station. In this case, the extension of the orbit of the satellite which can be usable for key distribution corresponds roughly to the interval $L \in [500,2000]$ km, corresponding to angles from the zenith in the interval $[0,80^\circ]$. The effective thickness of the atmosphere $\bar{h}$ is fixed here to $20$ km, for the considerations above.
	
	As introduced in Sec.\ref{freeintro}, we want to generalize the model proposed in \cite{vogel1,vogel2} to the just described case of a non-uniform link between the ground and a satellite. The computation follows the same steps and is described in \ref{free} and \ref{example}. First of all we need to evaluate Eqs.~(\ref{x02}), (\ref{varW}) and (\ref{varW2}) in order to compute the moments of the distributions of the elliptic beam parameters (Eq.~(\ref{paramv})). To do so, an integration along the propagation path must be performed (Eqs.~(\ref{dsturb}) and (\ref{dsscat})). Here we introduce the considerations of the previous paragraph, imposing that the parameters measuring the strength of the atmospheric effects are constant (greater than 0) inside the atmosphere and 0 outside. In particular we assume
	\begin{eqnarray}
	\label{nonuni}
	{\rm Down-links} \qquad &C_n^2(z)=C_n^2 \ \Theta(z-(L-h)) \nonumber \\
	&n_0(z)=n_0 \ \Theta(z-(L-h)) \nonumber \\
	{\rm Up-links} \qquad \quad &C_n^2(z)=C_n^2 \ \Theta(h-z) \nonumber \\
	&n_0(z)=n_0 \ \Theta(h-z) \ ,
	\end{eqnarray}
	where $C_n^2$ is the value of the refractive index structure constant and $n_0$ is the density of scattering particles. $\Theta(z)$ is the so-called Heaviside step-function, $z$ is the longitudinal coordinate, $L$ is the total length of the link and $h$ is the length traveled inside the atmosphere, as shown in Fig.~\ref{fig:atmo}. A down-link corresponds to the situation of satellite-to-ground communication, so the atmospheric effects kick-in only for $z>(L-h)$ (final section of the propagation), while for up-links it is limited to $z<h$. 
	We remark that some models for the altitude-dependence of the optical quantities, like $C_n^2$, are available in the literature \cite{hemmati,hufnagel,valley,lawson,frehlich}, but they are correct only in the geographical site and in the atmospheric conditions in which they had been experimentally extracted (more details in \ref{parameters}). Additional extinction losses due to back-scattering and absorption in the atmosphere are modeled by a parameter $\chi_{\rm  ext}$, as described in \ref{free}. Its value is adjusted from the analysis performed in \cite{jennewein1} based on the MODTRAN5 software \cite{modtran5}. In this model, the values of $C_n^2$ and $n_0$ completely describe the atmospheric conditions together with the thickness $\bar{h}$ and the extinction factor $\chi_{ext}$.
	
	Following the analysis of \ref{free} (in particular equations Eq.~(\ref{x02}), (\ref{varW}), (\ref{varW2})), we compute the first and second moments of the beam parameters in Eq.~(\ref{paramv}) for the link described in Eq.~(\ref{nonuni}). The distribution of the angle of orientation of the elliptical profile $\varphi_0$ is assumed uniform in $[0,\pi/2]$ as in \cite{vogel1,vogel2}. The mean value and variance of the beam centroid position are the same for $x$ and $y$ directions and equal to (Eq.~(\ref{x02}))
	
	\begin{equation}
	\label{distrXU}
	\langle x_0 \rangle=\langle y_0 \rangle=0 \ ,\qquad \langle x_0^2 \rangle=\langle y_0^2 \rangle= 0.419 \ \sigma_R^2 \ W_0^2 \ \Omega^{-\frac{7}{6}} \frac{h}{L} \ ,
	\end{equation}
	where the quantity $\sigma_R^2=1.23 \ C_n^2 \ k^{\frac{7}{6}} L^{\frac{11}{6}}$ is the so-called Rytov parameter and $\Omega=\frac{k W_0^2}{2 L}$ is the Fresnel number. The condition $\langle x_0 \rangle=0$ is achieved by proper pointing.
	The first two moments of the semi-axes of the ellipse squared, $W^2_i$ with $i=1,2$, are instead estimated from Eq.~(\ref{varW}) and (\ref{varW2})
	\begin{equation}
	\label{meanWU}
	\langle W_i^2 \rangle=\frac{W_0^2}{\Omega^2}\bigg( 1+\frac{\pi}{8} \ L \ n_0 \ W_0^2 \frac{h}{L}+2.6 \ \sigma_R^2 \ \Omega^{\frac{5}{6}} \frac{h}{L} \bigg)
	\end{equation}
	
	\begin{equation}
	\label{varWU}
	\langle \Delta W_i^2 \Delta W_j^2 \rangle=(2 \delta_{ij}-0.8) \ \frac{W_0^4}{\Omega^{\frac{19}{6}}} \bigg( 1+\frac{\pi}{8} \ L \ n_0 \ W_0^2 \frac{h}{L} \bigg) \sigma_R^2 \frac{h}{L}  \ .
	\end{equation}
	
	Similar expressions hold for down-links, for the beam centroid position
	
	\begin{equation}
	\label{distrXD}
	\langle x_0 \rangle=\langle y_0 \rangle=0 \qquad \langle x_0^2 \rangle=\langle y_0^2 \rangle= \alpha \ L \,
	\end{equation}
	
	and for the semi-axes of the elliptical profile
	
	\begin{equation}
	\label{meanWD}
	\langle W_i^2 \rangle=\frac{W_0^2}{\Omega^2}\bigg( 1+\frac{\pi}{24} \ L \ n_0 \ W_0^2 \bigg(\frac{h}{L}\bigg)^3+1.6 \ \sigma_R^2 \ \Omega^{\frac{5}{6}} \bigg(\frac{h}{L}\bigg)^\frac{8}{3} \bigg)
	\end{equation}
	
	\begin{equation}
	\label{varWD}
	\langle \Delta W_i^2 \Delta W_j^2 \rangle=(2 \delta_{ij}-0.8) \frac{3}{8}  \frac{W_0^4}{\Omega^{\frac{19}{6}}} \bigg( 1+\frac{\pi}{24} \ L \ n_0 \ W_0^2 \bigg(\frac{h}{L}\bigg)^3 \bigg) \sigma_R^2  \bigg(\frac{h}{L}\bigg)^\frac{8}{3} \ ,
	\end{equation}
	
	where $\alpha \sim 2 \ \mu$rad is the angular pointing error.
	
	There are two main differences between the expressions related to the up-link and down-link configurations. First, they depend on a different power of the ratio $\frac{h}{L}$. As $\frac{h}{L}\ll1$, we deduce, as expected, that the atmospheric effects are much stronger for up-links than for down-links. The phenomena involved here (beam deflection and broadening) are angular effects, whose contribution on the final size of the beam (and thus, on the losses of the channel) are proportional to the distance traveled after the "kick in" of the effect. For up-links, these effects happen very close to the transmitter, and then the beam broadens for hundreds of km before being detected. In the down-link scenario, instead, the beam travels in vacuum for the largest portion of the distance, and the atmospheric effects take place only at the end of the propagation, in the last tens of km before the receiver.
	The second difference resides in the origin of the fluctuations of the beam centroid position $x_0$. 
	For up-links, in fact, the deflections induced by the atmospheric effects are usually much stronger than the pointing error, which we neglect. For down-links, instead, at the top of the atmosphere the beam dimensions are already much larger than any turbulent inhomogeneity. In this case the induced beam wandering can be neglected and the pointing error becomes the main contribution.
	
	\begin{figure}[ht]
		\includegraphics[width=0.7\textwidth]{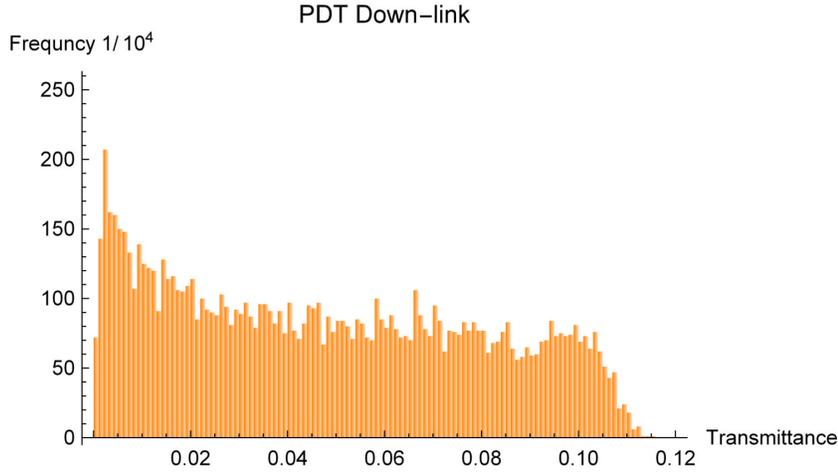}
		\centering
		\caption{\label{fig:dist} The Probability Distribution of the Transmittance (PDT) $\mathcal{P}(\eta)$ reconstructed by means of the method presented in Sec.\ref{modelresults} and \ref{free}. The situation under study is a down-link at high elevation angles ($L=500$ km) and the histogram has been obtained on the basis of 10000 events. The parameters of the setup are reported in \ref{parameters}.}
	\end{figure}
	
	\begin{figure}[ht]
		\includegraphics[width=0.65\textwidth]{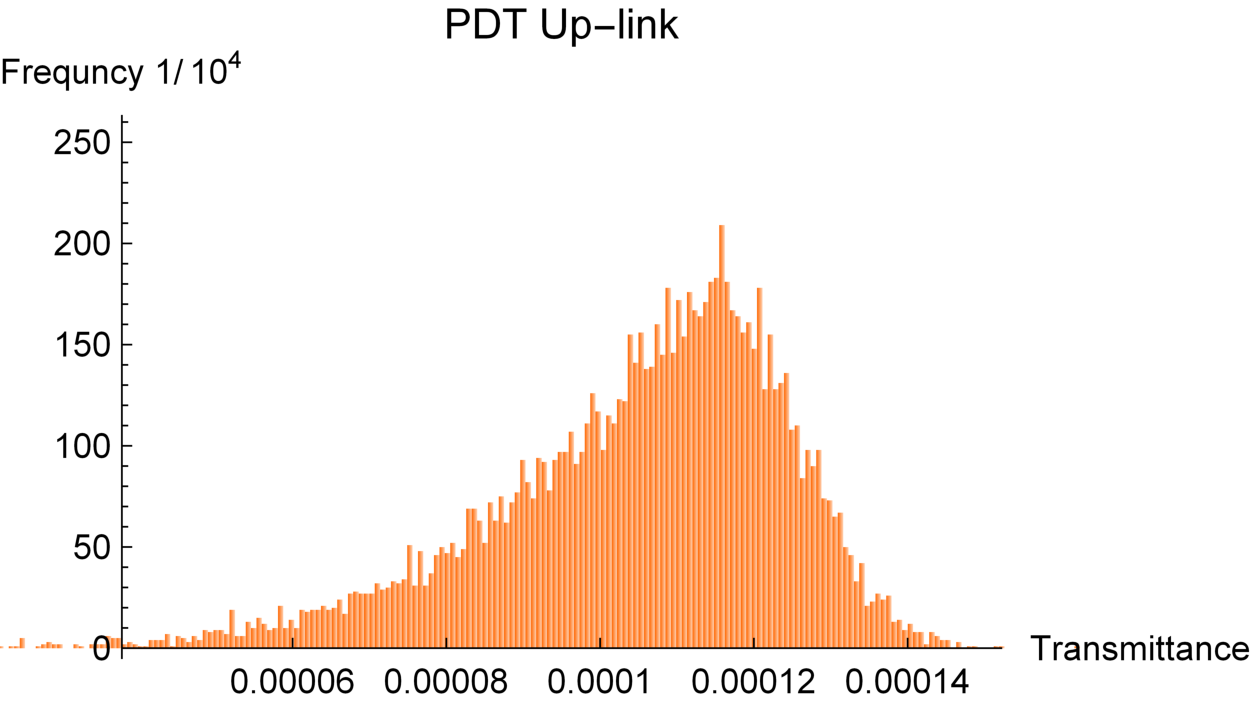}
		\centering
		\caption{\label{fig:distU} The Probability Distribution of the Transmittance (PDT) $\mathcal{P}(\eta)$ reconstructed by means of the method presented in Sec.\ref{modelresults} and \ref{free}. The situation under study is an up-link at high elevation angles ($L=500$ km) and the histogram has been obtained on the basis of 10000 events. The parameters of the setup are reported in \ref{parameters}.}
	\end{figure}
	
	The knowledge of the probability distribution of the elliptic beam parameters is then used to compute the PDT, through Eq.~(\ref{eta2}) and random sampling. Two examples are shown in Fig.~\ref{fig:dist} and Fig.~\ref{fig:distU} for a down-link and an up-link, respectively. The considerations of the previous paragraph can naturally be used to explain the difference in the shape of these two distributions. For down-links, especially at high elevation angles, like the case shown in Fig.~\ref{fig:dist}, the value of the beam width at the receiver is comparable to the wandering induced by pointing errors. This means that it can happen that the beam wanders completely off the receiving aperture, giving values of transmittance close to 0. In the up-link case, instead, the beam broadening gets the upper hand: the beam at the receiver is so large that the wandering induced by the atmosphere cannot change the total transmittance very much. It results in a rather narrow distribution, peaked at much lower values of transmittance with respect to the down-link case. 
	
\begin{figure}[ht]
		\includegraphics[width=0.75\textwidth]{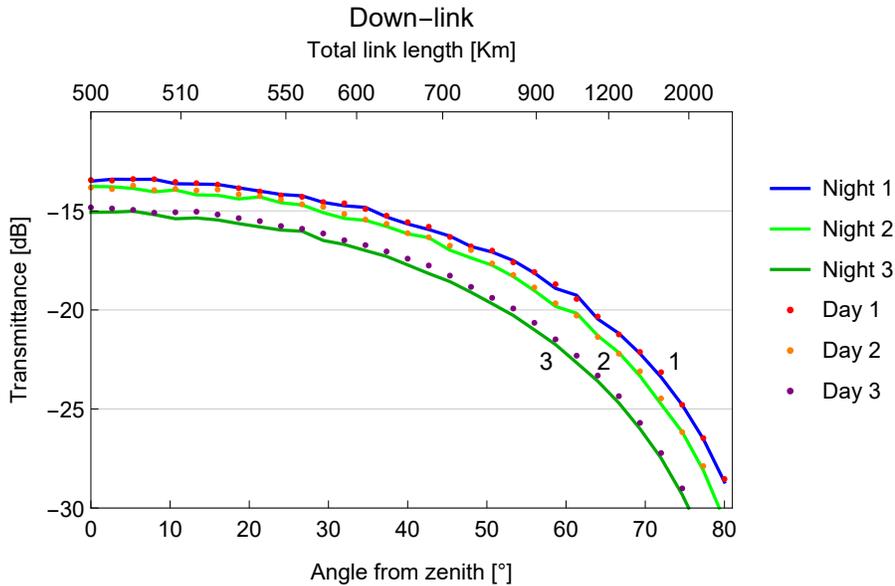}
		\centering
		\caption{\label{fig:TDown} Mean value of the Probability Distribution of the Transmittance (PDT) as a function of the zenith angle and total link length for the down-link configuration, under various weather conditions during night- and day-time. Situation 3 corresponds to worse weather conditions with respect to 2, that is in turn worse than 1. From a quantitative point of view, this means that the values of the parameters $C^2_n$ and $n_0$ grow going from 1 to 3. See Tab.~\ref{paraTab2} in \ref{parameters} for details about the choice of the parameters. From a qualitative point of view, they correspond to clear, slightly foggy and moderately foggy nights (Night 1-2-3) and to not windy, moderately windy and windy day (Day 1-2-3).
		Note that worse weather conditions generally correspond to higher extinction in the atmosphere. However, in order to highlight the contribution of the beam effects (broadening, wandering and shape distortion), we kept the value of $\chi_{\rm ext}$ fixed in this analysis, as well as in figure \ref{fig:TUp}. The non-uniformities are due to the finite statistics, every point corresponds to 1000 samples.}
	\end{figure}

\begin{figure}[ht]
		\includegraphics[width=0.75\textwidth]{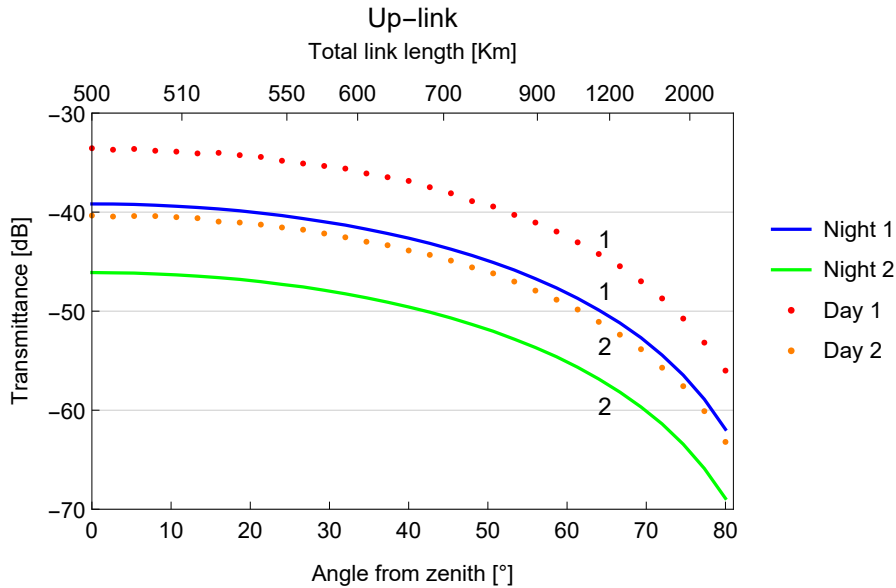}
		\centering
		\caption{\label{fig:TUp} Mean value of the Probability Distribution of the Transmittance (PDT) as a function of the zenith angle and total link length for the up-link configuration, under various weather conditions during night- and day-time. Same considerations as Fig.~\ref{fig:TDown} apply.}
	\end{figure}

	Now we want to study the expected loss introduced by the link as a function of the total link length. We show in figures \ref{fig:TDown} and \ref{fig:TUp} the mean value of the PDT as a function of the angle from the zenith and the total link length, for down-links and up-links, under different weather conditions. Every point in the graph has been obtained, just like in Fig.~\ref{fig:dist} and Fig.~\ref{fig:distU}, from 1000 samples of the parameters in Eq.~(\ref{paramv}) and using Eq.~(\ref{eta2}). The asymmetric nature of the PDT for some configurations of the link can make the use of the mean value partially misleading, however, the full PDT will be used in the next section to compute the secret key rates.
	
	The critical parameters here are, apart from the ones related to the atmospheric effects, the diameter of the sending and receiving telescopes and the signal wavelength. We chose $D_{\rm sat}=30$ cm for the orbiting one, $D_{\rm grnd}=1$ m for the ground station telescope and $\lambda=800$nm. These are demanding values, consistent with the Chinese mission \emph{Micius} (see \cite{Liao,pan2,pan3,pan4} for details). Further analysis are reported in Sec.\ref{cube}. 
	We notice here that the assumption of perfect Gaussian beams sent by the transmitter is not very realistic. Standard telescopes generate beams with intensity distributions rather close to a circular Gaussian profile but with some imperfections, introduced for example by the truncation at the border of the optical elements. The main downside is that such beams will exhibit larger intrinsic beam broadening due to diffraction. In our model this effect can be taken into account by adjusting the value of the initial beam waist $W_0$, in order to match the far-field divergence expected from the imperfect quasi-Gaussian beam.

	Our analysis confirms that, at least for the parameters chosen for the simulation, down-links are much preferable over up-links for quantum communication due to the smaller losses. However, up-links can still achieve losses below the threshold for the accomplishment of quantum communication tasks, QKD included. 
	Particularly interesting is the comparison between night- and day-time operation. During the day, the higher temperatures bring stronger wind and more active mixing between the different layers of the atmosphere, leading to more pronounced turbulence effects. However, on average, during clear days the moisture content of the lower atmosphere is smaller than at night, resulting in weaker beam spreading due to scattering particles. At night, instead, the lower temperature results, on one hand, in a less turbulent atmosphere and, on the other, in the formation of haze and mist. In this situation, the contribution of scattering over such particulate can be stronger than the turbulence-induced effects.
	
	Many different models for atmospheric channels and satellite-based links had already been proposed in the literature, due to the increasing interest in free-space optical communication. A comparison with them can highlight the strengths of the approach we reported in this section.
	Many feasibility studies \cite{aspel} rely on models that average the intensity over sufficiently long times, so that the only atmospheric effect is overall a broadening of the beam. This approach gives no information on the PDT of the channel, that can be useful in many instances (for example, to apply post-selection techniques). A different approach has been chosen by \cite{jennewein1}, based on convolution between the beam envelope and the time-averaged pointing errors and beam broadening, leading again to no information about the PDT. 
	A popular technique, that involves heavy numerical computations, is based on simulating the effect of the atmosphere by random phase screens regularly distributed along the propagation path in vacuum \cite{screens1,screens2,screens3}. Many theoretical works have been devoted to find the analytical probability distribution that better fits the experimentally measured transmittance of free-space optical links. Mainly used are the log-normal \cite{log1,log2}, Gamma-Gamma \cite{gammagamma} and Double Weibull \cite{weibull} distributions. Each of them appears to be more suitable depending on the strength of the turbulence, the length of the link and the configuration of the transmitting and receiving telescopes. On the contrary, the approach used here is a constructive method that allows to determine the PDT starting from the characteristics of the beam and the atmospheric conditions. It has been shown that a post-selection of the time-intervals with greater transmittance can help to increase the secret key rates \cite{arts1,arts2,jennewein3}: in this context, the ability of our approach to simulate not only the expected value of the transmittance, but its probability distribution too, may prove to be of great interest. Finally, we effectively take into account the contribution due to scattering particles, like fog or haze, making possible to model the effect of different weather conditions, a problem usually not addressed in previous works. It is particularly important during night-time operation, where a substantial amount of beam deformations can be imputed to scattering on moisture particles.

	\section{Performances of QKD implementations}
	\label{perf}
	
	The transmittance shown in Fig.~\ref{fig:TDown} and \ref{fig:TUp} can now be used to compute the expected secret key rates of a QKD protocol. In the following we analyze the performances of the BB-84 protocol \cite{bb84} with polarization encoding, implemented using either a true Single Photon (SP) source or Weak Coherent Pulses (WCPs). We use modern techniques to compute the secret key rates for SPs \cite{Toma2012} and WCPs with decoy states \cite{decoy3,decoy2,decoy4,decoy1}, taking into account finite-key effects. The key rates are averaged over the PDT computed for different link lengths and configurations
	
	\begin{equation}
	\bar{R}=\int_{0}^{1} R(\eta) \ \mathcal{P}(\eta) \ d\eta = \sum_{i=1}^{N_{\rm bins}} R(\eta_i) \ \mathcal{P}(\eta_i) \ .
	\end{equation}
	
	Here $\bar{R}$ is the averaged key rate, $R(\eta)$ the key rate at the specific value of the transmittance, $\mathcal{P}(\eta)$ is the PDT. The integral average is approximated dividing the range $[0,1]$ in $N_{\rm bins}$ bins, centered in $\eta_i$ for $i=1, \ N_{\rm bins}$, and taking the weighted sum of the rates. $\mathcal{P}(\eta_i)$ is estimated through random sampling, as pointed out in Sec.~\ref{modelresults}. The expressions for the key rates $R(\eta)$ for the different implementations are given in \ref{rates}, see Eq.~(\ref{ratesingle}) for SPs and Eq.~(\ref{ratewcp}) for WCPs.
	
	The biggest source of noise in free-space optical links is represented by environmental light entering in the receiver telescope together with the signal photons. Simple models to estimate the amount of stray light \cite{bonato,miao} are given in \ref{envphot} for down-links and up-links. In the following analysis we consider the number of stray photons to be independent of the position of the satellite. Particular situations concerning light pollution, like the presence of a city close to the ground station, may require a more specific model for low elevation angles.
	
\begin{figure}[ht]
		\includegraphics[width=0.4\textwidth]{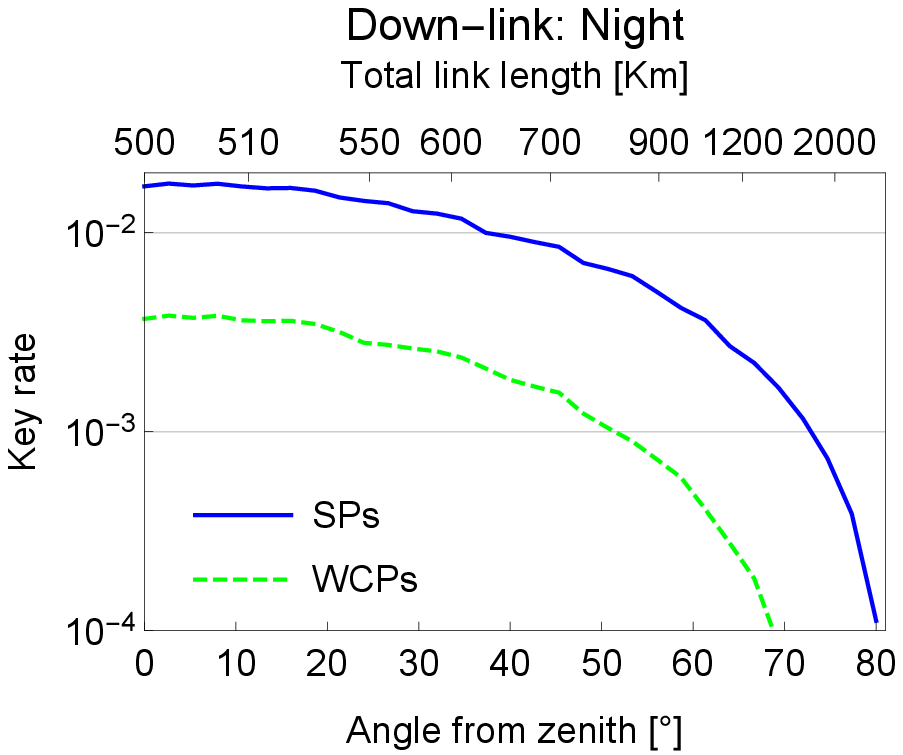}
		\includegraphics[width=0.4\textwidth]{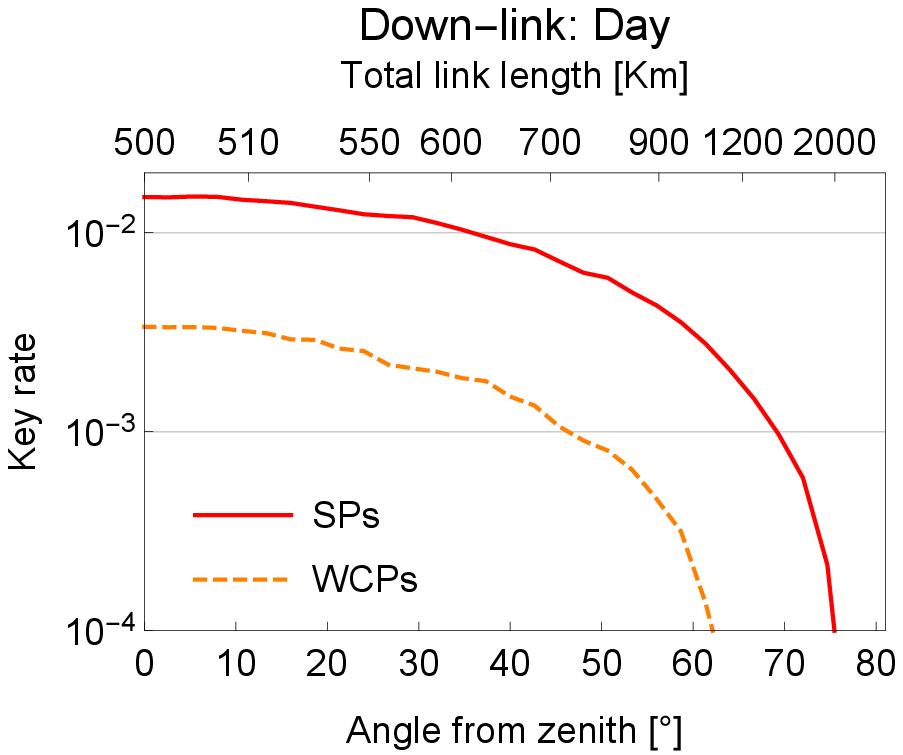}
		\centering
		\caption{\label{fig:Down} The expected key rate generated by the BB-84 protocol with SP and WCP implementations is reported as a function of the zenith angle and the total link length, for a down-link. We assume here good weather conditions, corresponding to situation 1 in Fig.~\ref{fig:TDown}.}
	\end{figure}

\begin{figure}[ht]
		\includegraphics[width=0.4\textwidth]{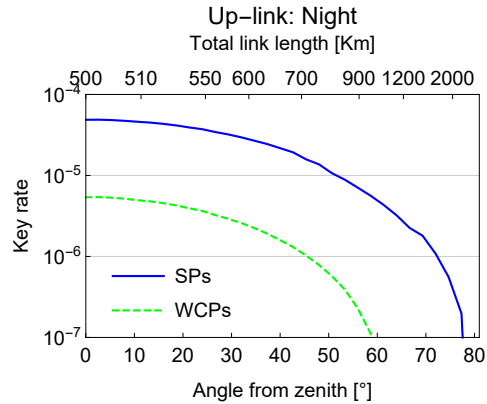}
		\centering
		\caption{\label{fig:Up} The expected key rate generated by the BB-84 protocol with SP and WCP implementations is reported as a function of the zenith angle and the total link length, for an up-link. We assume here good weather conditions, corresponding to situation 1 in Fig.~\ref{fig:TDown}.}
	\end{figure}

	The secret key rate resulting from a down-link and an up-link are reported in Fig.~\ref{fig:Down} and Fig.~\ref{fig:Up}, for both night-time and day-time operation, under good weather conditions, corresponding to situation 1 in Fig.~\ref{fig:TDown}. In the following we set the block sizes at $10^6$ for SPs and at $10^8$ for WCPs in down-link. This difference is justified by the higher repetition rates obtainable by modern WCP sources with respect to (still under development) true SP sources. Consider that the total link duration is around 300 s, corresponding to the complete passage of a LEO satellite over the ground station. In this time span, assuming a repetition rate of 10 MHz for SP sources and 1 GHz for WCP sources, several blocks of the size specified above can be exchanged in the down-link configuration. Due to the higher losses encountered in an up-link, the block size is lowered to $10^5$ for SPs and at $10^7$ for WCPs.
	
	At night it is possible to establish a non-zero key rate in down-link during the whole passage of the satellite in the SP implementation. Using WCPs, instead, the key rate drops to 0 when the satellite is around $20^\circ$ over the horizon. In the daytime, instead, due to the stronger background light, the key rate vanishes at higher elevation angles, even considering improved spatial, spectral and temporal filtering (refer to Tab.~\ref{paraTab3} in \ref{parameters}).
	
	Up-links have poorer performances due to higher losses, but we are still able to distill a secret key with non-zero rates during the night, with slightly improved filtering (Tab.~\ref{paraTab3} in \ref{parameters}). The SP implementation reaches almost the same range (in elevation angle) as the down-link configuration, while the difference with WCPs is greater because of the smaller block size. For day-time operation the stronger background light makes the quantum bit error rate too high and the key rate vanishes, therefore we omit the corresponding graph. We stress that here (we refer to \ref{envphot} for details) we did not consider artificial light pollution. So these results reliably simulate only ground stations which are isolated and far from big cities. 
	
	Note that the finite key effects can be very detrimental when the number of exchanged signals becomes too small. Particular attention must be payed when up-links are considered. In order to reproduce the results reported in Fig.~\ref{fig:Up}, the block length used in the security analysis is of the same order of magnitude of the number of signals exchanged during the whole passage of the satellite. This means that all the signals exchanged in a QKD session are processed in a single block in this case.
	
	\section{Cube-sat performance analysis}
	\label{cube}
	
\begin{figure}[ht]
		\includegraphics[width=0.75\textwidth]{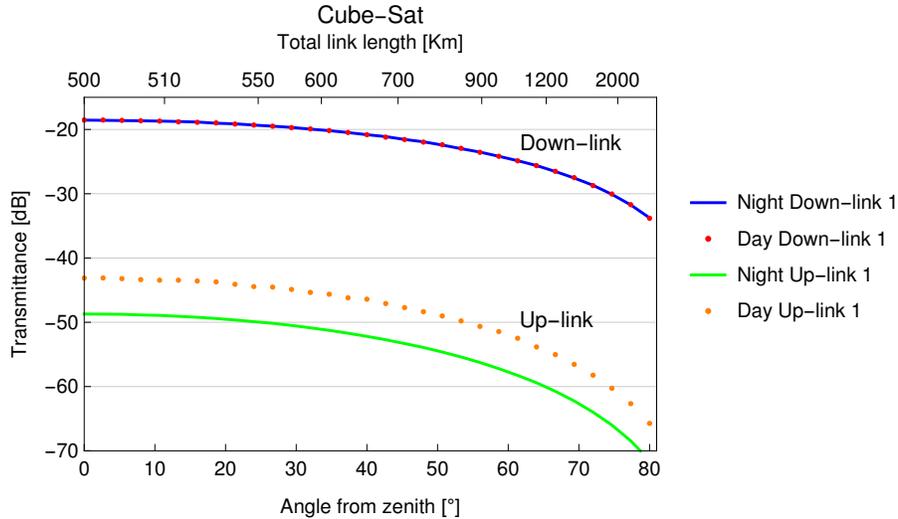}
		\centering
		\caption{\label{fig:Cube} Mean value of the Probability Distribution of the Transmittance (PDT) as a function of the zenith angle and total link length for up-link and down-link configurations, using a Cube-sat with a 10 cm telescope. The weather conditions correspond to situation 1 in Fig.~\ref{fig:TDown} and Fig.~\ref{fig:TUp}.}
	\end{figure}

	The simulations reported in Sec.~\ref{modelresults} and Sec.~\ref{perf} assume a quite demanding value of the optical aperture of the orbiting telescope. It is compatible with the {\it Micius} satellite \cite{Liao,pan2,pan3,pan4}, operated by the {\it Chinese Academy of Science}, as part of the {\it Quantum Experiments at Space Scale} (QUESS) research project. The complexity and high cost of the mission make the use of such big satellites unfeasible for the establishment of a world-wide quantum communication network. 
	Many recent proposals foresee the use of nano-satellites (e.g., \emph{Cube-sats} \cite{cubeG1,cubeG2,cubeG3}) for QKD implementation \cite{Bedington,cube1,cube2,cube3,cube4}. The possibility to deploy many of such satellites in a single mission, or to share the vector with other payloads, lowers considerably the launch cost of these devices. They must be loaded with much smaller optics, of diameter $\leq 10$ cm, even if bigger aperture can be achieved with the use of deployable optics. When used as transmitter, in the down-link configuration, the smaller aperture creates beams with much higher intrinsic divergence than the case studied in Sec.~\ref{modelresults}. In the up-link configuration, instead, smaller transmittance is due to the smaller collecting area. We show in Fig.~\ref{fig:Cube} the results of the link simulation for down-links and up-links, in good weather conditions.   
	
	We see that the effect of the smaller optics diameter amounts to a difference in transmittance of about 5 dB for down-link and to 10 dB for up-links. Even though this result favors the down-link configuration even more, we have to take into account that a smaller aperture will collect not only less signal light, but also less stray light. The resulting Quantum Bit Error Rate for up-links, then, will be almost independent of the diameter of the receiving telescope. 
	
\begin{figure}[ht]
		\includegraphics[width=0.40\textwidth]{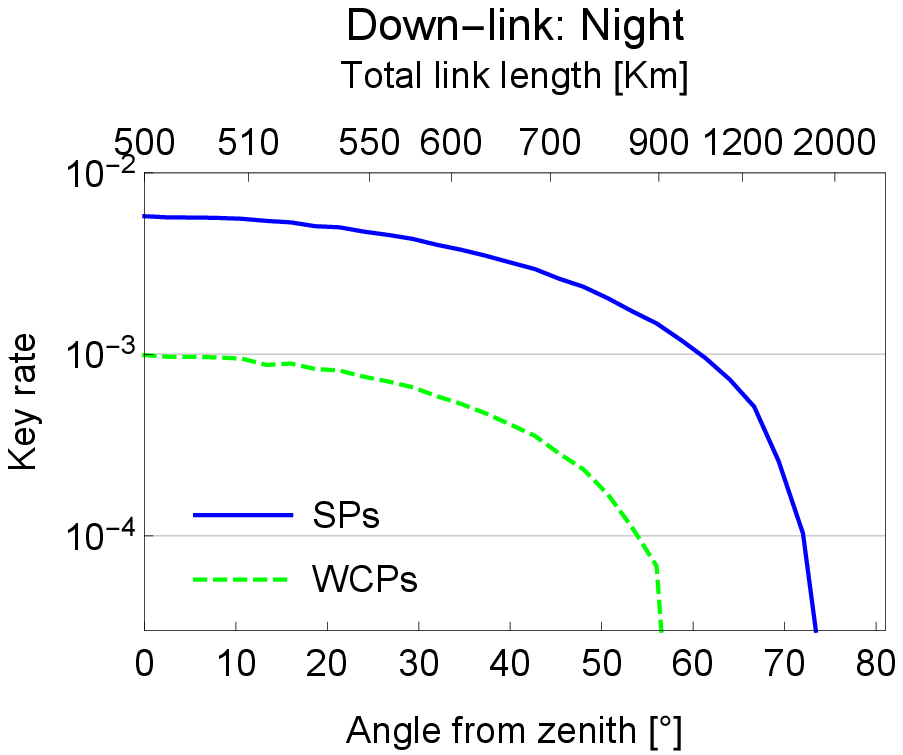}
		\includegraphics[width=0.40\textwidth]{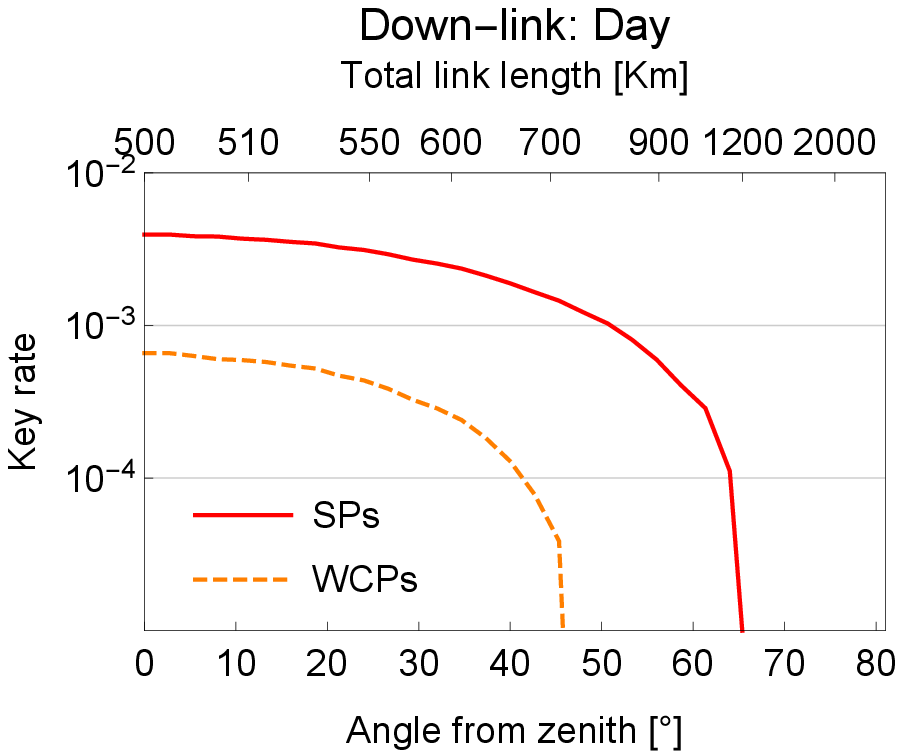}
		\centering
		\caption{\label{fig:DownC} The expected key rate generated by the BB-84 protocol with SP and WCP implementations is reported as a function of the zenith angle and the total link length, for a down-link using a Cube-Sat. We assume here good weather conditions, corresponding to situation 1 in Fig.~\ref{fig:TDown}}
	\end{figure}

\begin{figure}[ht]
		\includegraphics[width=0.40\textwidth]{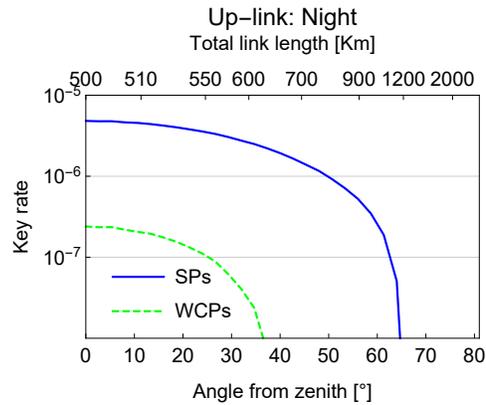}
		\centering
		\caption{\label{fig:UpC} The expected key rate generated by the BB-84 protocol with SP and WCP implementations is reported as a function of the zenith angle and the total link length, for an up-link using a Cube-Sat. We assume here good weather conditions, corresponding to situation 1 in Fig.~\ref{fig:TDown}}
	\end{figure}

	The key rates achievable for nano-satellites in the down-link and up-link configurations are reported in Fig.~\ref{fig:DownC} and Fig.~\ref{fig:UpC}. As expected, the range of angles over which a non-zero key rate can be exchanged shrinks with respect to the case of Sec.~\ref{perf}. We point out that we kept the block length fixed at the values reported in \ref{perf} even if, especially in the up-link configuration, that number of signals can't be exchanged in a single transit of the Cube-sat.
	
	\section{Conclusion}
	\label{discussion}
	
	We provide a general and fundamental model to simulate the losses introduced by a satellite-based optical link, useful for feasibility and performance analysis of future free-space QKD experiments. The ability to precisely evaluate the contribution due to different weather conditions will be crucial in many situations. 
	The geographical sites with better performances can be more precisely mapped, in order to optimize the structure of future global quantum networks \cite{network1,network2,network3,network4,network5}. 
	Through the use of this model, the data from meteorological predictions can directly be linked to the performances achievable by the QKD link, allowing more accurate statistical studies of the number of operative days per year. 
	The characterization of the transmittance of the channel has then be used to evaluate the performances of the link in terms of achievable secret key rates. We focused on two implementations of the BB-84 cryptographic protocol, using single photons and weak coherent pulses. The noise expected in interesting real-life scenarios, during night-time and day-time, has been modeled and taken into account. We also pointed out the importance of finite-key effects, which can be very detrimental due to the short duration of the link between ground station and satellite. 
	The simulations confirm that long-distance quantum communications can be achieved not only using medium-sized satellites, like the Chinese \emph{Micius}, but also nano-satellites, allowing to considerably cut the cost of a space-based global quantum network. 
	Ultimately, such links are expected to be integrated with a repeater-based quantum network on the ground, to complement it and enhance its performances when long distances need to be bridged. The analysis of such a configuration and the optimization of its topology and structure are still under study and represent a crucial milestone towards the realization of the dreamt quantum internet.

	\ack{This project has received funding from the European Union's Horizon 2020 research and innovation programme under the Marie Sk\l{}odowska-Curie grant agreement No. 675662.}

\clearpage
\newpage

\appendix
\section{Free-space links with turbulence and scatterers}
\label{free}


In this section we summarize the analysis of atmospheric optical channels proposed in \cite{vogel1,vogel2}. We will discuss the background, show the main steps of the derivation and recap some results, that will be used as starting point for the simulations described in Sec.~\ref{modelresults}. 

We start from the introduction to the problem given in Sec.~\ref{intro}. 
The solution of the paraxial wave equation, with phase-approximation using the Huygens-Kirchhoff method \cite{phaseappr}, can be written in the following way
\begin{eqnarray}
\label{env}
u(\brho,L)&=& \int_{\mathbb{R}} d^2 \brho' u_0(\brho') G_0(\brho,\brho';L,0) \nonumber \\
&\times& {\rm exp}[i S(\brho,\brho';z,z')] \ .
\end{eqnarray}
Since the losses due to back-scattering and absorption can't be included in the paraxial approximation of the Helmholtz equation, we treat them phenomenologically multiplying the beam envelope $u(\brho,L)$ by $\sqrt{\chi_{\rm ext}}$. The extinction factor $\chi_{\rm ext} \in [0,1]$ accounts for absorption and back-scattering losses and can be considered as a non-fluctuating quantity (see \cite{vogel2}). In Eq.~(\ref{env}) $u_0(\brho')$ is the Gaussian envelope at the transmitter plane ($z=0$, $z$ is the longitudinal coordinate) \begin{equation}
\label{u0}
u_0(\brho)=\sqrt{\frac{2}{\pi W_0^2}} \ {\rm exp}\bigg[ -\frac{1}{W_0^2}|\brho|^2-\frac{i k}{2F} |\brho|^2 \bigg] \ ,
\end{equation} 
with $W_0$ the beam spot radius at the transmitter, $k$ the optical wavenumber and $F$ the focal length of the beam.
$G_0$ is a Gaussian integral kernel
\begin{eqnarray}
\label{G0}
G_0(\brho,\brho':z,z')=\frac{k}{2 \pi i (z-z')} \ {\rm exp}\bigg[\frac{i k |\brho-\brho'|^2}{2(z-z')}\bigg]  
\end{eqnarray}
while $S$ contains all the atmospheric effects
\begin{equation}
\label{S}
S(\brho,\brho';z,z')=\frac{k}{2} \int_{z'}^{z} d \zeta \ \delta \varepsilon\bigg(\brho\frac{\zeta-z'}{z-z'}+\brho'\frac{z-\zeta}{z-z'},\zeta\bigg) \ .
\end{equation}
Here $S(\brho,\brho';z,z')$ gives the phase contribution due to inhomogeneities of the relative permittivity of the air $\delta \epsilon(\brho'',\xi)$ from $z'$ to $z$. Note that $\delta \varepsilon$ can be separated in two contributions, related to turbulence and scattering
\begin{equation}
\label{epsfact}
\delta \varepsilon=\delta \varepsilon_{\rm turb}+\delta \varepsilon_{\rm scat} \ .
\end{equation}
Assuming the two contributions to be statistically independent, the same factorization holds for the permittivity fluctuation spectrum
\begin{equation}
\label{phifact}
\Phi_\varepsilon ({\bf K})=\Phi_\varepsilon^{\rm turb} ({\bf K}) + \Phi_\varepsilon^{\rm scat} ({\bf K}) \ ,
\end{equation}
defined as the Fourier transform of the correlation function of $\delta \varepsilon({\bf r})$
\begin{equation}
\langle \delta \varepsilon({\bf r}_1) \delta \varepsilon({\bf r}_2) \rangle= \int d^3 {\bf K} \ \Phi_\varepsilon ({\bf K}) \ {\rm exp}[i {\bf K}\cdot ({\bf r}_1-{\bf r}_2)] \ .
\end{equation}
In the previous equations ${\bf K}$ denotes the momentum and is a 3-dimensional vector.
The Markov approximation (that in our case corresponds to assume delta-correlation in the $z$ direction) simplify this expression \cite{turb43,turb45}
\begin{equation}
\langle \delta \varepsilon({\bf r}_1) \delta \varepsilon({\bf r}_2) \rangle= 2 \pi \delta(z_1-z_2) \int d^2 {\bf k} \ \Phi_\varepsilon ({\bf k}) \ {\rm exp}[i {\bf k}\cdot ({\bf \brho}_1-{\bf \brho}_2)] 
\end{equation}
where ${\bf k}$ represents the momentum in the plane transverse to the propagation direction. $\brho_1$ and $\brho_2$ are the components of the vectors ${\bf r}_1$ and ${\bf r}_2$ in the transversal plane, while $\delta(z)$ is the Dirac-delta.
The Kolmogorov model allows us to write the turbulence-related part of the relative permittivity fluctuation spectrum as \cite{kolmogorov,turb43,turb44,turb45}
\begin{equation}
\label{kolm}
\Phi_\varepsilon^{\rm turb}({\bf k})=0.132 \ C_n^2 \ |{\bf k}|^{-\frac{11}{3}} \ .
\end{equation}
The refractive index structure constant $C_n^2$ characterizes the strength of turbulence in the optical domain and is an important parameter of the model.
The scattering term in Eq.~(\ref{phifact}) can be approximated as a Gaussian function 
\cite{scatt19,scatt20,scatt21,scatt22,scatt23}
\begin{equation}
\label{phiscat}
\Phi_\varepsilon^{\rm scat}({\bf k})=\frac{n_0 \zeta_0^4}{8 \pi k^2} \ {\rm exp} \big[-\zeta_0^2|{\bf k}|^2\big] \ ,
\end{equation}
with $\zeta_0$ correlation length of the fluctuations due to scattering particles. Here $n_0$ is the mean number of scatterers per unit volume and represents the main parameter in describing the strength of the scattering contribution.

Now we want to use these ingredients to calculate the probability distribution of the parameters in the elliptic beam approximation, introduced in Sec.~\ref{intro}
\begin{equation}
\label{paramv2}
{\bf v} = (x_0,y_0,W_1, W_2, \varphi_0) \ .
\end{equation}
First of all we define normalized variables from the ellipse semi-axes 
\begin{equation}
\label{thetai}
\Theta_i=\ln\bigg(\frac{W_i^2}{W_0^2}\bigg) \quad i=1,2 \ ,
\end{equation} 
where $W_0$ is the beam spot radius at the transmitter.
Now we assume that, in the case of uniform turbulence and scatterers density, the probability distribution of $x_0,y_0,\Theta_1,\Theta_2$ is Gaussian, while the angle of orientation $\varphi_0$ is uniformly distributed in $[0,\pi/2]$ (see Appendix of \cite{vogel2} for details). The mean value and the variance of these distributions can be analytically computed. We recall the main steps of the derivation in the following paragraphs.  

Starting from the beam centroid position $(x_0,y_0)$, we can choose the reference frame such that $\langle x_0 \rangle = \langle y_0 \rangle = 0$ and \cite{turb43,andrews}
\begin{equation}
\label{x02}
\langle x_0^2 \rangle=\langle y_0^2 \rangle=\int_{\mathbb{R}^4} d^2 \boldsymbol{\rho}_1 d^2 \boldsymbol{\rho}_2 x_1 x_2 \Gamma_4(\boldsymbol{\rho}_1,\boldsymbol{\rho}_2;L) \ .
\end{equation}
Here $\Gamma_4(\boldsymbol{\rho}_1,\boldsymbol{\rho}_2;z)=\langle u^*(\boldsymbol{ \rho}_1,z) u(\boldsymbol{ \rho}_1,z) u^*(\boldsymbol{ \rho}_2,z) u(\boldsymbol{ \rho}_1,z) \rangle$ is the fourth-order field correlation function. 

The means and covariances of the squared ellipse semi-axes $W_i^2$ have the following form (see Appendix of \cite{vogel1} for details)
\begin{eqnarray}
\label{varW}
&&\langle W_{1/2}^2 \rangle = 4\bigg[ \int _{\mathbb{R}^2} d^2 \boldsymbol{\rho} \ x^2 \ \Gamma_2(\boldsymbol{\rho};L) - \langle x_0^2 \rangle \bigg] \ , \\
&&\langle \Delta W_{i}^2 \Delta W_{j}^2 \rangle = - 8 \biggl\{ 2 \bigg( \int_{\mathbb{R}^2}d^2 \boldsymbol{\rho} \ x^2 \ \Gamma_2(\boldsymbol{\rho};L)\bigg)^2 \nonumber \\
&& - \int_{\mathbb{R}^4} d^2 \boldsymbol{\rho}_1 \ d^2 \boldsymbol{\rho}_2 \ [x_1^2 x_2^2 (4 \delta_{ij}-1) - x_1^2 y_2^2 (4\delta_{ij} -3)] \nonumber \\
\label{varW2}
&& \times \Gamma_4(\boldsymbol{\rho}_1,\boldsymbol{\rho}_2;L) \bigg\} - 16 [4 \delta_{ij}-1]\langle x_0^2 \rangle^2 \ , 
\end{eqnarray} 
where the second-order field correlation function $\Gamma_2(\boldsymbol{\rho};z)=\langle u^*(\boldsymbol{ \rho},z) u(\boldsymbol{ \rho},z)\rangle$ has been used.

The next step is the calculation of the field correlation function, for which we use the expression of the beam envelope given in Eq.~(\ref{env}). We report the calculations only for $\Gamma_2(\brho;L)$, the equivalent but more cumbersome expressions for $\Gamma_4(\brho_1,\brho_2;L)$ can be found in \cite{vogel2}, Appendix B. 
Substituting Eq.~(\ref{env}) in the definition of $\Gamma_2(\brho;z)$ yields
\begin{eqnarray}
\label{gamma2}
\Gamma_2(\brho&&;L) =  \int_{\mathbb{R}^4} d^2 \ \brho_1' d^2 \brho_2' u_0(\brho_1') u_0^*(\brho_2') G_0(\brho,\brho_1';L,0) \nonumber \\
&&\times G_0^*(\brho,\brho_2';L,0) \ {\rm exp} \bigg[ - \frac{1}{2} \mathcal{D}_S(0,\brho_1'-\brho_2') \bigg] \ , 
\end{eqnarray}
with the last term embodying the phase fluctuations due to the atmosphere (remember the definition of $S(\brho,\brho';z,z')$ in Eq.~(\ref{S}))
\begin{eqnarray}
\mathcal{D}_S&&(\brho_k-\brho_l,\brho_k'-\brho_l') \nonumber \\
&&= \bigg\langle [S(\brho_k,\brho_k';z,z')-S(\boldsymbol{\rho}_l,\brho_l';z,z')]^2 \bigg\rangle \ .
\end{eqnarray} 
Substituting Eq.~(\ref{S}) and exploiting again the Markov approximation, the factorization in Eq.~(\ref{epsfact}) and (\ref{phifact}) can be carried over
\begin{eqnarray}
\mathcal{D}_S = \mathcal{D}_S^{\rm turb}+\mathcal{D}_S^{\rm scat} \ .
\end{eqnarray}
We can now introduce the models for the permittivity fluctuations spectrum related to turbulence (Eq.~(\ref{kolm})) and scatterers (Eq.~(\ref{phiscat})), obtaining 
\begin{eqnarray}
\label{dsturb}
&&\mathcal{D}_S^{\rm turb}(\brho,\brho') =2.95 \ k^{2} \ L \int_0^{1} d \xi \ C_n^2(\xi) \ |\brho \xi + \brho'(1-\xi)|^{\frac{5}{3}} \\
\label{dsscat}
&&\mathcal{D}_S^{\rm scat}(\brho,\brho') = \frac{\pi}{8} L \int_0^1 d \xi \ n_0(\xi) \ |\brho \xi + \brho'(1-\xi)|^{2}
\end{eqnarray}
where we introduced the rescaled longitudinal coordinate $\xi\in [0,1]$, where $\xi=1$ corresponds to $z=L$. We allowed for a dependence on longitudinal coordinate in $C_n^2(\xi)$ and $n_0(\xi)$ for later use. We recall the definition of the so-called Rytov parameter $\sigma_R^2=1.23 \ C_n^2 \ k^{\frac{7}{6}} L^{\frac{11}{6}}$.
Substituting in Eq.~(\ref{gamma2}) the definition of the Gaussian envelope $u_0(\brho)$ (Eq.~(\ref{u0})) and the integral kernel $G_0(\brho,\brho':L,0)$ (Eq.~(\ref{G0})), the second-order field correlation function reads 
\begin{eqnarray}
\label{gamma22}
\Gamma_2(\brho;L)=\frac{\Omega^2}{\pi^2 W_0^4} \int_{\mathbb{R}^2} && d^2 \brho' \ {\rm e}^{-\frac{\alpha}{2 W_0^2}|\brho'|^2 - 2 i \frac{\Omega}{W_0^2} \brho \cdot \brho'} \nonumber \\
{\rm exp} \bigg[-\frac{1}{2} \mathcal{D}_S^{turb}(0,\brho') \bigg] && {\rm exp} \bigg[-\frac{1}{2} \mathcal{D}_S^{scat}(0,\brho') \bigg] \ .
\end{eqnarray}
Here $\alpha=1+\Omega^2 \big( 1-\frac{L}{F} \big)^2$ with the Fresnel number defined as $\Omega=\frac{k W_0^2}{2 L}$. 

Integrating Eq.~(\ref{gamma22}) (and the equivalent one for $\Gamma_4$) and then Eqs.~(\ref{varW}), we obtain the first and second moments of the probability distribution of $W_i^2$. Then, the moments for the variables $\Theta_i$ are easily obtained from Eq.~(\ref{thetai}).

We consider now the transmittance, defined in Eq.~(\ref{eta}), of an elliptic beam impinging on a circular aperture of radius $a$. It can be written as
\begin{eqnarray}
\label{eta2}
\eta(&x_0,y_0,W_1, W_2, \varphi_0) =\\
&=\frac{2 \ \chi_{\rm ext}}{\pi W_1 W_2} \int_{0}^{a} d \rho \int_{0}^{2\pi} d\theta \ {\rm e}^{-2 A_1(\rho \ {\rm cos} \theta - \rho_0)^2} \nonumber\\ 
&\times {\rm e}^{-2 A_2 \rho^2 {\rm sin}^2 \theta } {\rm e}^{-2 A_3(\rho \ {\rm cos} \theta - \rho_0)r \ \sin\theta} \ , \nonumber
\end{eqnarray}
with
\begin{eqnarray}
A_1&=&\bigg(\frac{\cos^2(\varphi_0-\theta_0)}{W_1^2}+\frac{\sin^2(\varphi_0-\theta_0)}{W_2^2}\bigg) \\
A_2&=&\bigg(\frac{\sin^2(\varphi_0-\theta_0)}{W_1^2}+\frac{\cos^2(\varphi_0-\theta_0)}{W_2^2}\bigg) \nonumber \\
A_3&=&\bigg(\frac{1}{W_1^2}-\frac{1}{W_2^2}\bigg) \ \sin \ 2(\varphi_0-\theta_0) \ .\nonumber
\end{eqnarray}
In the previous equations $(\rho,\theta)$ are the integration variables in the area of the circular aperture, while $(x_0,y_0)=(\rho_0 \ \cos\theta_0,\rho_0 \ \sin\theta_0)$ is the beam-centroid position. 

The Probability Distribution of the Transmittance (PDT) is then easily reconstructed. Extract at random $M$ 5-tuples of values for $(x_0,y_0,\Theta_1, \Theta_2, \varphi_0)$, according to the correct probability distribution. Compute first the values of the ellipse semi-axes $W_i$ from $\Theta_i$ and then the value of the transmittance for every tuple. Collect the statistics in an histogram and compute statistical estimators (e.g., the median). Two examples of the simulated PDT are shown in Sec.~\ref{modelresults} of the main text (Fig.~\ref{fig:dist} and Fig.~\ref{fig:distU}).

\section{Application of the model to a satellite-based link}
\label{example}

In this section we are going to apply the model described in the previous section to a satellite-based link, as described in Sec.~\ref{modelresults} of the main text. We will discuss some details about the calculations involved and show an example of how to proceed with the integration of the expressions in \ref{free}. In particular, we will focus on the first term of the quantity $\langle W_{1/2}^2 \rangle$ defined in equation Eq.~(\ref{varW}), which only contains the second order correlation function
$\Gamma_2(\brho;L)$. 
The computations involving the integration of the fourth order correlation function $\Gamma_4(\brho_1,\brho_2;L)$ are much more cumbersome and will not be reported here.

Inserting Eq.~(\ref{gamma22}) in the first term of Eq.~(\ref{varW}) we obtain the following integration, where all the quantities are defined in the previous section \ref{free}

\begin{eqnarray}
\label{intgamma2}
\int_{\mathbb{R}^2} d^2 \brho \ x^2 \ \Gamma_2(\brho;L)&&=\frac{\Omega^2}{\pi^2 W_0^4} \int_{\mathbb{R}^4}  d^2 \brho \ d^2 \brho' \ x^2 \ {\rm e}^{-\frac{\alpha}{2 W_0^2}|\brho'|^2 - 2 i \frac{\Omega}{W_0^2} \brho \cdot \brho'} \nonumber \\
&&{\rm exp} \bigg[-\frac{1}{2} \mathcal{D}_S^{\rm turb}(0,\brho') \bigg]  {\rm exp} \bigg[-\frac{1}{2} \mathcal{D}_S^{\rm scat}(0,\brho') \bigg] \ .
\end{eqnarray}

We assume that the beam is focused ($F=L$) so that $\alpha=1$. First of all we can compute the terms $\mathcal{D}_S^{\rm turb}(0,\brho')$ and $\mathcal{D}_S^{\rm scat}(0,\brho')$ defined in Eq.~(\ref{dsturb}) and (\ref{dsscat}) 

\begin{eqnarray}
\label{dsturb0}
&&\mathcal{D}_S^{\rm turb}(0,\brho') =2.95 \ k^{2} \ L \int_0^{1} d \xi \ C_n^2(\xi) \ |\brho'(1-\xi)|^{\frac{5}{3}} \\
\label{dsscat0}
&&\mathcal{D}_S^{\rm scat}(0,\brho')= \frac{\pi}{8} L \int_0^1 d \xi \ n_0(\xi) \ |\brho'(1-\xi)|^{2} \ .
\end{eqnarray} 

In the rescaled longitudinal coordinate $\xi$, the conditions in Eq.~(\ref{nonuni}) in the main text become

\begin{eqnarray}
\label{nonunixi}
{\rm Down-links} \qquad &C_n^2(\xi)=C_n^2 \ \Theta(\xi-(1-h/L)) \nonumber \\
&n_0(\xi)=n_0 \ \Theta(\xi-(1-h/L)) \nonumber \\
{\rm Up-links} \qquad \quad &C_n^2(\xi)=C_n^2 \ \Theta(h/L-\xi) \nonumber \\
&n_0(\xi)=n_0 \ \Theta(h/L-\xi) \ .
\end{eqnarray}

Inserting Eq.~(\ref{nonunixi}) (we consider down-link in this example) into Eq.~(\ref{dsturb0}) and (\ref{dsscat0}) we can solve the integration and obtain

\begin{eqnarray}
\label{dsturb00}
&&\mathcal{D}_S^{\rm turb}(0,\brho') =2.95 \ k^{2} \ L \ |\brho'|^{\frac{5}{3}} C_n^2 \int_{1-h/L}^{1} d \xi \  \ (1-\xi)^{\frac{5}{3}} \nonumber \\
&&= 2.4 \ \sigma^2_R \ k^{5/6} \ L^{-5/6} \ |\brho'|^{\frac{5}{3}} \bigg[ \frac{3}{8} \bigg( \frac{h}{L} \bigg)^{8/3} \bigg]\\
\label{dsscat00}
&&\mathcal{D}_S^{\rm scat}(0,\brho')= \frac{\pi}{8} L  n_0 |\brho'|^{2} \int_{1-h/L}^{1} d \xi \  \ |\brho'(1-\xi)|^{2} \nonumber \\
&&= \frac{\pi}{8} L n_0 |\brho'|^{2}  \bigg[ \frac{1}{3} \bigg( \frac{h}{L} \bigg)^{3} \bigg]
\end{eqnarray}

where $\sigma^2_R$ has been defined in Sec.~\ref{modelresults}. From this passage we clearly see where the dependency on $\frac{h}{L}$ in Eq.~(\ref{distrXU}) to Eq.~(\ref{varWD}) originates from. When we introduce Eq.~(\ref{dsturb00}) and Eq.~(\ref{dsscat00}) in Eq.~(\ref{intgamma2}), we recognize that it only contains Gaussian integrals of the form

\begin{equation}
\int_{-\infty}^{\infty} d x \ x^c \ \exp [a \ x^2 + i \ b \ x] \ ,
\end{equation}
with $c=\{0,2\}$ and that can be readily solved. The only exception is the turbulence term, which contains $|\brho'|^{\frac{5}{3}}$. We can simplify the computation introducing the approximation \cite{vogel2,andrews} $|\brho'/W_0|^{\frac{5}{3}} \simeq |\brho'/W_0|^{2}$. Then, one just has to solve the multiple Gaussian integrals and insert it in Eq.~(\ref{varW}) to obtain the value of $\langle W_{1/2}^2 \rangle$ for down-links, as in Eq.~(\ref{meanWD}). Similar techniques can be used to compute all the other moments of the beam variables.

Eq.~(\ref{distrXU}), (\ref{meanWU}), (\ref{distrXD}) and (\ref{meanWD}) have been computed specifically for the problem at hand, the non-uniform link described at the beginning of Sec.~\ref{modelresults}. Eq.~(\ref{varWU}) and (\ref{varWD}), on the other hand, have been deduced from the equivalent results obtained in \cite{vogel2} for a uniform link. 
We see that in Eq.~(\ref{distrXU}), (\ref{meanWU}), (\ref{distrXD}) and (\ref{meanWD}) the corrections due to the non-uniformity of the link (of the form $A (h/L)^\beta$, where $A$ is a constant and $\beta=\{1,8/3,3\}$) act like multiplicative factors on the parameters $\sigma^2_R$ and $n_0$. So, we started from the calculation of the quantity $\langle \Delta W_{i}^2 \Delta W_{j}^2 \rangle$ in \cite{vogel2} and attached the multiplicative corrections found above, in order to obtain Eq.~(\ref{varWU}) and (\ref{varWD}).
This inconsistency should not be considered too detrimental regarding the reliability of the model. We checked through the simulation that the mean value and the shape of the PDT are not very sensitive to variations of the value of the quantities in Eq.~(\ref{varWU}) and (\ref{varWD}), as the interplay between beam wandering (Eq.~(\ref{distrXU}) and (\ref{distrXD})) and beam spreading (Eq.~(\ref{meanWU}) and (\ref{meanWD})) is much more significant in this context.
Finally, we point out that the computation of the quantities in Eq.~(\ref{distrXU}), (\ref{meanWU}), (\ref{distrXD}) and (\ref{meanWD}) have been carried out without the introduction of the weak turbulence approximation used in \cite{vogel1,vogel2}. For Eq.~(\ref{varWU}) and (\ref{varWD}), instead, we used the results obtained in \cite{vogel2} in the weak turbulence regime, which we verified to be still valid in the case of satellite-based links.

\section{Error model and environmental photons}
\label{envphot}

In a free-space link, environmental photons are usually the most important source of noise. In this section we summarize the analysis of \cite{bonato,miao} regarding the amount of environmental photons that hit the detector for down-links and up-links, that we use to calculate the expected QBER. We suppose that an accurate time synchronization had been operated between sender and receiver, in order to tag the photons and perform a time filtering on the incoming signal. On top of that, wavelength filtering is applied to further reduce the amount of detected noisy photons.

For up-links, we only consider the case of night-time operation. If the ground station site has a low level of light pollution, the biggest fraction of environmental photons comes from the Sunlight reflected first by the Moon and then by the Earth \cite{bonato}
\begin{equation}
N_{\rm night}^{\rm up}=A_E A_M R_M^2 a^2 \frac{\Omega_{\rm fov}}{d^2_{EM}} B_f \ \Delta t \ H_{\rm sun} \ .
\end{equation}
Here $A_M$ and $R_M$ are the albedo and the radius of the Moon, while $A_E$ is the albedo of the Earth and $d_{EM}$ is the Earth-Moon distance. $H_{\rm sun}$ is the solar spectral irradiance in ${\rm photons} \ {\rm s}^{-1} {\rm nm}^{-1} {\rm m}^{-2}$ at the wavelength of interest. $\Omega_{\rm fov}$ and $a$ are angular field of view and radius of the receiving telescope. $B_f$ is the width of the spectral filtering and $\Delta t$ is the detection time-window. We assumed Lambertian diffusion on the Moon and the Earth.

For down-links, the evaluation of the background photons is strongly site-dependent. The power received by the telescope can be expressed as follows \cite{miao}
\begin{equation}
\label{p}
P_b = H_b \Omega _{\rm fov} \pi a^2 B_f \ .
\end{equation}
The parameter $H_b$ is the total brightness of the sky background and it depends on the hour of the day and the weather conditions. From Eq.~(\ref{p}) we derive the number of photons per time window 
\begin{equation}
N^{\rm down}=\frac{H_b}{h \nu} \Omega_{\rm fov} \pi a^2 B_f \ \Delta t \ ,
\end{equation}
where $h$ is the Planck constant and $\nu$ is the frequency of the background photons (after filtering). Typical values of the brightness of the sky are $H_b=10^{-3} \ W \ {\rm m}^{-2} \ {\rm sr} \ \mu {\rm m}$ during a full-Moon night and $H_b=1 \ W \ {\rm m}^{-2} \ {\rm sr} \ \mu {\rm m}$ for a clear sky in day-time. This analysis assumes that neither the Moon during the night nor the Sun during the day are included in the field of view of the collecting aperture.

The Quantum bit error rate (QBER) is computed assuming the noisy photons to be completely unpolarized
\begin{equation}
QBER=Q_{0} + \frac{1}{2} \frac{N_{\rm noise}}{N_{\rm noise} + N_{\rm sig}} \ .
\end{equation} 

Here $Q_{0}$ corresponds to the error rate associated with depolarization in the encoding degree of freedom or imperfection of the preparation or detection stage leading to incorrect state discrimination. We chose a conservative value of $Q_{0}=2 \%$. $N_{\rm noise}$ and $N_{\rm sig}$ are, respectively, the expected number of photons per time window associated to noise and signal.
As expected, the number of collected environmental photons are proportional to the area of the receiving aperture, but so is the intensity of the signal. To reduce the noise and at the same time raise the signal to noise ratio, we can act on $\Omega_{\rm fov}$, $B_f$ and $\Delta t$. Reducing the field of view involves a better pointing and tracking system, while a very good time synchronization allows the use of short time windows.

\section{Rates for BB-84 with single photons and Weak Coherent Pulses}
\label{rates}

We report here the expression of the secret key rates we used in the performance study of section \ref{perf}. The set-up is the usual one for QKD: two parties, A and B, are connected through a completely insecure quantum channel and an authenticated classical channel. After many uses of the links, their goal is to share an identical key, which is secret regardless of the attack strategy that an hypothetical eavesdropper could implement. 
For the single-photon implementation of the BB-84 protocol (using, e.g., polarization encoding), party A sends qubits in the basis $X=\{\ket{0},\ket{1}\}$ or $Z=\{\ket{+},\ket{-}\}$ at random, with $\ket{\pm}=(\ket{0}\pm \ket{1})/\sqrt{2}$. B measures the received qubits in the bases $X$ or $Z$, at random. The results of \cite{Toma2012} state that a secret key of length $l$ can be shared, if  

\begin{eqnarray}
\label{ratesingle}
&l \leq n (q-h_2(Q_{\rm tol}+\mu))-{\rm leak}_{EC}-\alpha(\varepsilon_{\rm sec},\varepsilon_{\rm cor}) \nonumber \\ 
&\alpha(\varepsilon_{\rm sec},\varepsilon_{\rm cor})=\log_2\frac{2}{\varepsilon_{\rm sec}^2 \varepsilon_{\rm cor}} \ \ \mu=\sqrt{\frac{n+k}{nk}\frac{k+1}{k} \ln\frac{2}{\varepsilon_{\rm sec}}} \ ,
\end{eqnarray}

out of $n$ successfully exchanged single photon signals, where the function $h_2$ denotes the binary entropy. Here $q$ is a parameter describing the preparation quality of the initial states of the signal sent by A. In the qubit case it is connected to the maximum fidelity allowed between states prepared in the $X$ and $Z$ bases. In a perfect implementation of the BB-84 protocol, like the one considered here, the two bases are mutually unbiased, for which the maximum $q=1$ is achieved. $Q_{\rm tol}$ is the channel error tolerance and $k$ is the number of bits of the raw key used for parameter estimation. The achievable key rate is obtained by maximizing over these two parameters. The term ${\rm leak}_{EC}$ gives the amount of information in bits that the parties had to exchange during the error correction phase. The desired security and correctness thresholds are specified by the parameters $\varepsilon_{\rm sec}$ and $\varepsilon_{\rm cor}$.

An alternative protocol based on decoy states \cite{decoy3,decoy4} is used when the source emits Weak Coherent Pulses instead of real single photons. We follow the analysis of \cite{decoy2,decoy1}, where two decoy states are used. The bases used for the encoding are $X$ and $Z$ as in the single photon implementation. A secret key of length $l$ can be extracted, with

\begin{equation}
\label{ratewcp}
l \leq s_{X,0}+s_{X,1}(1-h_2(\phi_X)) \nonumber 
- {\rm leak}_{EC}-6 \ \log_2\frac{21}{\varepsilon_{\rm sec}}-\log_2 \frac{2}{\varepsilon_{\rm cor}} \ .
\end{equation}

$s_{X,0}$ and $s_{X,1}$ represent the number of bits in the raw key generated by vacuum events and single photon events, respectively. $\phi_X$ instead is the phase error rate measured in the channel during parameter estimation. The subscript $X$ means that these estimations are valid for the events in which both A and B chose the basis $X$ and they include the corrections due to finite key effects (for the actual expressions we refer to \cite{decoy1}). In this case the maximization is over the portion of signals used for parameter estimation, the intensity of the signal and decoy states and the probability of sending different intensities. 

In both cases, the key rates are obtained taking the ratio between the length in bit of the final secret key $l$ and total number of signals sent $n$.

\section{Choice of parameters for the satellite-based link}
\label{parameters}

In this section we show the values oft the parameters utilized throughout the paper and we discuss about their pertinence. They are reported in Tab.~\ref{paraTab1},  Tab.~\ref{paraTab2} and Tab.~\ref{paraTab3}, together with a brief explanatory description, where necessary. More detailed explanations about particular parameters are in the remainder of this section.

\begin{table}[h!]
	\centering
	\begin{tabular}{ | c | c | c | } 
		\hline
		Parameter & Value & Brief description \\ 
		\hline
		$W_0$ & 15 cm, 50 cm & down-links, up-links \\ 
		\hline
		$W_0$ & 5 cm, 50 cm & CubeSat down-links, up-links \\ 
		\hline
		$a$ & 50 cm, 15 cm & down-links, up-links \\ 
		\hline
		$a$ & 50 cm, 5 cm & CubeSat down-links, up-links \\ 
		\hline
		$\lambda$ & 785 nm & Wavelength of the signal light \\ 
		\hline
		$\beta$ & 0.7 & Parameter in $\chi_{\rm ext}(\theta)$  \\ 
		\hline
		$\alpha$ & $1.2 \ 10^{-6} \ {\rm rad} $ & Pointing error \\ 
		\hline
		$\eta_{\rm det}$ & 0.5 & Detector efficiency \\
		\hline
		$T_{\rm opt}$ & 0.8 & Transmittance of the optical system \\
		\hline
	\end{tabular}
	\caption{Parameters related to the optical and technical properties of the link.}
	\label{paraTab1}
\end{table}

\begin{table}[h!]
	\centering
	\begin{tabular}{ | c | c | c | } 
		\hline
		Parameter & Value & Brief description \\ 
		\hline
		$h$ & 20 km & Atmosphere thickness \\ 
		\hline
		$L$ & 500 km & Minimum altitude (zenith) \\ 
		\hline
		$C_n^2$ & $1.12 \ 10^{-16} \ {\rm m}^{-2/3}$ & Night-time, condition 1 \\ 
		\hline
		$C_n^2$ & $1.64 \ 10^{-16} \ {\rm m}^{-2/3}$ & Day-time, condition 1 \\ 
		\hline
		$C_n^2$ & $5.50 \ 10^{-16} \ {\rm m}^{-2/3}$ & Night-time, condition 2 \\ 
		\hline
		$C_n^2$ & $8.00 \ 10^{-16} \ {\rm m}^{-2/3}$ & Day-time, condition 2 \\
		\hline
		$C_n^2$ & $1.10 \ 10^{-15} \ {\rm m}^{-2/3}$ & Night-time, condition 3 \\ 
		\hline
		$C_n^2$ & $1.60 \ 10^{-15} \ {\rm m}^{-2/3}$ & Day-time, condition 3 \\
		\hline
		$n_0$ & 0.61 ${\rm m}^{-3}$ & Night-time, condition 1 \\ 
		\hline
		$n_0$ & 0.01 ${\rm m}^{-3}$ & Day-time, condition 1 \\ 
		\hline
		$n_0$ & 3.00 ${\rm m}^{-3}$ & Night-time, condition 2 \\ 
		\hline
		$n_0$ & 0.05 ${\rm m}^{-3}$ & Day-time, condition 2 \\ 
		\hline
		$n_0$ & 6.10 ${\rm m}^{-3}$ & Night-time, condition 3 \\ 
		\hline
		$n_0$ & 0.10 ${\rm m}^{-3}$ & Day-time, condition 3 \\ 
		\hline
	\end{tabular}
	\caption{Parameters related to the atmospheric weather conditions.}
	\label{paraTab2}
\end{table}

\begin{table}[h!]
	\centering
	\begin{tabular}{ | c | c | c | } 
		\hline
		Parameter & Value & Brief description \\
		\hline
		Sky brightness $H_b$ & $1.5 \ 10^{-6}\ {\rm W} \ {\rm m}^{-2} \ {\rm sr}^{-1} {\rm nm}^{-1}$ & Night, clear sky, full Moon \cite{miao} \\  
		\hline
		Sky brightness $H_b$ & $1.5 \ 10^{-3}\ {\rm W} \ {\rm m}^{-2} \ {\rm sr}^{-1} {\rm nm}^{-1}$ & Day, clear sky \cite{miao} \\  
		\hline
		Field of view $\Omega_{\rm fov}$ & $(100 \ 10^{-6})^2$ sr &  Night-time down-link \\  
		\hline
		Field of view $\Omega_{\rm fov}$ & $(10 \ 10^{-6})^2$ sr &  Day-time down-link \\  
		\hline
		Field of view $\Omega_{\rm fov}$ & $(10 \ 10^{-6})^2$ sr &  Night-time up-link \\  
		\hline
		Time-window $\Delta t$ & 1 ns & Night- and day-time \\  
		\hline
		Spectral filter width $B_f$ & 1 nm & Night-time down-link \\
		\hline 
		Spectral filter width $B_f$ & 0.2 nm & Day-time down-link \\ 
		\hline
		Spectral filter width $B_f$ & 1 nm & Night-time up-link \\ 
		\hline
		$H_{\rm sun}$ & $4.610 \ 10^{18} \ {\rm phot} \ {\rm s}^{-1} {\rm nm}^{-1} {\rm m}^{-2}$ & Solar spectral irradiance \\ 
		\hline
		$A_{e}$ & 0.300 & Earth's albedo\\ 
		\hline
		$A_{m}$ & 0.136 & Moon's albedo\\ 
		\hline
	    $R_M$ & $1.737 \ 10^6$ m & Moon's radius\\ 
		\hline
		$d_{EM}$ & $3.600 \ 10^8$ m & Earth-Moon distance\\ 
		\hline
	\end{tabular}
	\caption{Parameters related to stray photons and environmental light.}
	\label{paraTab3}
\end{table}

The parameters $C_n^2$, $n_0$ and $h$ should in general be fixed by fitting the experimental data. However, in order to have a predictive model, we want to estimate these parameters in a reasonable way. First of all, in order to estimate the effective thickness of the atmosphere, we start from the variation of density of the air as a function of the altitude. We chose $h=20$ km, as a layer around the Earth with this thickness contains on average 95\% of the total mass of the atmosphere. As already stated in the main text, some models for the altitude dependence of the refractive index structure constant $C_n^2$ are available in the literature \cite{hufnagel,valley,lawson,frehlich}. The widely used parametric fit due to Hufnagel and Valley \cite{hufnagel,valley} reliably replicates the behaviour of $C_n^2$ in mid-latitude climate

\begin{eqnarray}
\label{intCn2}
C_n^2(z)&=5.94 \ 10^{-53} \bigg(\frac{v}{27}\bigg)^2 z^{10} \ {\rm exp}[-z/1000]+ \nonumber \\ 
&+ 2.7 \ 10^{-16} \ {\rm exp}[z/1500] + A \ {\rm exp}[z/100] \ .
\end{eqnarray}

Here $z$ is the altitude coordinate, $v$ is a parameter related to high-altitude wind speed and $A$ describes the relative strength of the turbulence near the ground level. Typical values are $A=1.7 \ 10^{-14} \ {\rm m}^{-2/3}$ and $v=21 \ {\rm m}/{\rm s}$, although $v=57 {\rm m}/{\rm s}$ is sometimes used for stronger wind conditions. The value of $C_n^2$ inside the atmosphere in our model is estimated by the integral average of this function in $[0,\infty]$, rescaled by the fixed thickness $h$

\begin{equation}
C_n^2=\frac{1}{h}\int_{0}^{\infty} C_n^2(z) \ dz \ .
\end{equation}

The parameter $v$ is kept fixed to the recommended value of $21 \ {\rm m}/{\rm s}$. $A$ is chosen to match the values of $C_n^2(0)$ measured in \cite{vogel2}, $A_n=1.10 \ 10^{-14} \ {\rm m}^{-2/3}$ at night and $A_d=2.75 \ 10^{-14} \ {\rm m}^{-2/3}$ during the day. Through Eq.~(\ref{intCn2}), the first corresponds to $C_n^2=1.12 \ 10^{-16} \ {\rm m}^{-2/3}$ and the latter to $C_n^2=1.64 \ 10^{-16} \ {\rm m}^{-2/3}$.

The scattering particles described by the density $n_0$ mainly consist of water droplets, so, in order to estimate the value of $n_0$, we start from the profile of the water vapour content in the atmosphere. The absolute humidity vertical profile $\tau(z)$ in the range $[0,10 \ {\rm km}]$ can be written as a double exponential \cite{tomasi,tomasi2}

\begin{eqnarray}
\tau(z)&=\tau(0) \ {\rm exp}[-z/H_1] \qquad \qquad \qquad  \ \ {\rm for} \  0 \le z \le 5 \ {\rm km} \\ \nonumber
&=\tau(H_1) \ {\rm exp}[-(z-5 \ {\rm km})/H_2] \qquad {\rm for} \  5 \ {\rm km} \le z \le 10 \ {\rm km}
\end{eqnarray}

with the two scale heights $H_1$ and $H_2$. The contribution of the region with $z>10 \ {\rm km}$ is rather low and we neglect it here. The parameters $H_1$ and $H_1$ can on average vary in the range $[1.53,2.8]$ and $[1.19,1.82]$, respectively, depending on the geographical position and the season. We choose in the following the values stated in the U. S. Standard Atmosphere (1962) \cite{stdAtmoUS}, $H_1=2.243$ and $H_2=1.414$. We obtain a rescaling factor $\omega$ in the same way as we did in the previous case

\begin{equation}
\omega=\frac{1}{h \ \tau(0)} \int_{0}^{10 \ {\rm km}} \tau(z) \ dz \ .
\end{equation}

Then, the value of the parameter $n_0^*$ in our case is obtained multiplying by the factor $\omega$ the value found in \cite{vogel2} for night- and day-time, $n_0^{*}=\omega \ n_0$. For the given values of the scale heights $\omega \simeq 0.107$. 

The extinction factor $\chi_{\rm ext}(\theta)$ varies as a function of the elevation angle in the following way

\begin{equation}
\chi_{\rm ext}(\theta)={\rm exp}\big[ -\beta \ {\rm sec}(\theta)  \big]
\end{equation}

The value of the parameter $\beta$ reported in Tab.~\ref{paraTab1} has been chosen to match the amount of extinction used in \cite{jennewein1}, based on the MODTRAN5 software \cite{modtran5}.

\newpage
\bibliographystyle{unsrt}
\bibliography{satelliteNew}

\end{document}